\documentclass[preprint,12pt]{elsarticle}

\usepackage{amssymb}
\usepackage[utf8]{inputenc}
\usepackage[normalem]{ulem}
\usepackage{graphicx}
\usepackage{multirow}
\usepackage{amsmath,amsfonts}
\usepackage{amsthm}
\usepackage{mathrsfs}
\usepackage[title]{appendix}
\biboptions{sort&compress}
\usepackage{xcolor}
\usepackage{textcomp}
\usepackage{manyfoot}
\usepackage{booktabs}
\usepackage{algorithm}
\usepackage{algorithmicx}
\usepackage{algpseudocode}
\usepackage{listings}
\usepackage{hyperref}

\journal{Advances in Water Resources}

\begin{document}

\title{Computationally Efficient and Error Aware Surrogate Construction for Numerical Solutions of  Subsurface Flow Through Porous Media}

\begin{frontmatter}

\author[inst1,inst4]{Aleksei G. Sorokin}
\author[inst1,inst2]{Aleksandra Pachalieva}
\author[inst2]{Daniel O'Malley}
\author[inst1]{James M. Hyman}
\author[inst4]{Fred J. Hickernell}
\author[inst3]{Nicolas W. Hengartner}

%\affiliation[inst1]{organization={Los Alamos National Laboratory}, Center for Nonlinear Studies
%            addressline={Bikini Atoll Road}, 
%            city={Los Alamos},
%            postcode={87545}, 
%            state={New Mexico},
%            country={United States}}

\affiliation[inst1]{organization=
{Center for Non-Linear Studies, Los Alamos National Laboratory},              city={Los Alamos},
            postcode={87545}, 
            state={New Mexico},
            country={United States}}

\affiliation[inst2]{organization=
{Energy and Natural Resources Security Group (EES-16), Earth and Environmental Sciences Division, Los Alamos National Laboratory},
            city={Los Alamos},
            postcode={87545}, 
            state={New Mexico},
            country={United States}}

\affiliation[inst3]{organization=
{Theoretical Biology and Biophysics Group (T-6), Theoretical Division, Los Alamos National Laboratory},
            city={Los Alamos},
            postcode={87545}, 
            state={New Mexico},
            country={United States}}

\affiliation[inst4]{organization={Illinois Institute of Technology}, Department of Applied Mathematics
            addressline={10 W 35th Street}, 
            city={Chicago},
            postcode={60616}, 
            state={Illinois},
            country={United States}}

\begin{abstract}
    % 1. The question we are addressing is 
    Limiting the injection rate to restrict the pressure below a threshold at a critical location can be an important goal of simulations that model the subsurface pressure between injection and extraction wells.   
    The pressure is approximated by the solution of Darcy's partial differential equation for a given permeability field. The subsurface permeability is modeled as a random field since it is known only up to statistical properties. This induces uncertainty in the computed pressure.
    Solving the partial differential equation for an ensemble of random permeability simulations enables estimating a probability distribution for the pressure at the critical location.   
    % 2. The bottleneck is 
    These simulations are computationally expensive, and practitioners often need rapid online guidance for real-time pressure management.  
    % 3. The Methods we use to address this problem are (what we did)
    An ensemble of numerical partial differential equation solutions is used to construct a Gaussian process regression model that can quickly predict the pressure at the critical location as a function of the extraction rate and permeability realization. The Gaussian process surrogate analyzes the ensemble of numerical pressure solutions at the critical location as noisy observations of the true pressure solution, enabling robust inference using the conditional Gaussian process distribution. 
    
    % 4. Our principal results are 
    Our first novel contribution is to identify a sampling methodology for the random environment and matching kernel technology for which fitting the Gaussian process regression model scales as $\mathcal{O}(n \log n)$ instead of the typical $\mathcal{O}(n^3)$ rate in the number of samples $n$ used to fit the surrogate.  
    The surrogate model allows almost instantaneous predictions for the pressure at the critical location as a function of the extraction rate and permeability realization. Our second contribution is a novel algorithm to calibrate the uncertainty in the surrogate model to the discrepancy between the true pressure solution of Darcy's equation and the numerical solution. 
    % 5. What these results mean
    Although our method is derived for building a surrogate for the solution of Darcy's equation with a random permeability field, the framework broadly applies to solutions of other partial differential equations with random coefficients.
\end{abstract}


% \begin{highlights}
% \item Developed an error-aware surrogate model for the random pressure at a critical location.  
% \item Presented a computationally efficient method of fitting the surrogate model, which enables scaling to a large number of samples. 
% \end{highlights}


\end{frontmatter}

%\newpage

%\tableofcontents

%\AlekseiNote{
%\section*{TODO}
%\begin{itemize}
%    \item Better formatting
%    \begin{itemize}
%        \item Graphical abstract
%        \item Research highlights
%        \item better keywords
%        \item PACS and MSC codes
%        \item remove unnecessary packages and commands
%    \end{itemize}
%    \item Open source QuasiGaussianProcesses.jl
%    \item Reproducible code in DPFEHM package? 
%    \item Update reference to GaussianRandomFields.jl with JORS paper
%\end{itemize}}

%\notes{problem formulation Darcy's law}

%\notes{generating an ensemble of permeability fields }

%\notes{Karhunen-Loève expansion of the permeability field }

%\notes{Solving the PDE and the DPFEHM package}

%\notes{The error bounds}

%\notes{creating the GPR surrogate model}

%\notes{overview of the algorithm}

%\notes{numerical simulations}

%\notes{state main results}

%\newpage 

\section{Introduction}
% The problem we are addressing is 

Large-scale geologic CO$_2$ sequestration (GCS) is a promising technology to combat global warming and climate change caused by rapidly rising atmospheric carbon dioxide levels. A well-chosen CO$_2$ sequestration reservoir would keep 99\% of the carbon dioxide sequestered for over 1000 years \cite{metz2005ipcc}. The key to minimizing the risk of leakage and induced seismicity is to develop a reservoir pressure management strategy for choosing the reservoir sites. These sites need to be robust against failure while minimizing cost. 
%Mishandling the pressure management increases the risk of triggering leakage, inducing seismicity, and creating faults and fractures \cite{buscheck2011combining,cihan2015optimal,harp2017development}.   

Successful pressure management strategies for GCS are essential to prevent overpressurization in the subsurface caused by resource extraction and carbon sequestration \cite{viswanathan2008development,benson2008co2,birkholzer2009basin,stauffer2011greening,middleton2012cross,gholami2021leakage}. Failure to accomplish this goal can lead to induced seismicity \cite{majer2007induced,zoback2012managing,keranen2014sharp,mcnamara2015earthquake}, leakage of sequestered resources such as CO$_2$ into the atmosphere \cite{buscheck2011combining,cihan2015optimal,harp2017development,chen2018geologic,chen2019characterization} and along abandoned wellbores \cite{pruess2008co,watson2009evaluation,nordbotten2009model,carey2010experimental,huerta2013experimental,jordan2015response,harp2016reduced,yonkofski2019risk,lackey2019managing,mehana2022reduced}, and potential contamination of water aquifers \cite{keating2010impact,little2010potential,trautz2013effect,navarre2013elucidating,keating2016reduced,bacon2016modeling,xiao2020chemical}.
These events erode public distrust, increase economic costs and financial risk, create obstacles to deployment of future projects \cite{bielicki2016leakage,gholami2021leakage}, and even project cancellation\cite{palmgren2004initial,curry2005aware,miller2007public,wilson2008regulating,court2012promising,tcvetkov2019public,whitmarsh2019framing}.

GCS operations will benefit from a better understanding and risk analysis tools to support pressure management in subsurface flow fields. To address these issues, a complex physics models must be solved with sufficient fidelity and enough realizations to reduce the inherent uncertainties in the heterogeneous subsurface, as well as in the GCS site characterization and operations \cite{ben2006info,omalley2015bayesian,chen2020reducing,vasylkivska2021nrap}. 
Yet existing pressure management models are often costly to fit and do not account for the discretization error in numerical simulations. To overcome these challenges, our surrogate model for the pressure exploits strategically selected sampling locations and a matching covariance kernel to enable fast model fitting and evaluation. Moreover, the discretization error in our numerical simulations is systematically encoded into the model through the use of noisy observation and noise variance calibration.

When modeling subsurface flow between injection and extraction wells, there are scenarios where it is important to ensure that the pressure at a critical location is below a given threshold value. To simplify the presentation of our results, consider the situation where fluid is injected down a single well at a fixed rate into a heterogeneous porous media and the increased subsurface pressure pushing other fluids out at a single extraction well. To calculate the
subsurface pressure one must solve Darcy's partial differential equation.

% \begin{figure}[tb]
%     \centering
%     \includegraphics[width=.8\textwidth]{subsurface_flow_diagram_bw.png}
%     \caption{Fluid is injected at a fixed rate into porous media with uncertain properties. This has the potential to increase the pressure throughout the subsurface. Some fluid is extracted at a different location to decrease the pressure. The extraction rate is chosen so the pressure at a critical location is below a given threshold with high probability. \AlekseiNote{You can edit the black and white version at \url{https://docs.google.com/drawings/d/1UUHAImvugXrZCy1-fZjMX88JLBVx1vbHVrxm5QBGlm4/edit?usp=sharing} and the color version at \url{https://docs.google.com/drawings/d/17yS8qJBmMisQtxoWEKIKgD_j9Q7gTTO25ehVEnk2XVk/edit?usp=sharing}}}
%     \label{fig:subsurface_flow_diagram}
% \end{figure}

%The subsurface pressure can be modeled by Darcy's partial differential equation when the subsurface permeability field is known. However, the permeability field is generally unknown and is characterized as a random field described by a correlated probability distribution. Estimating the probability distribution for the pressure at the critical location requires analyzing an ensemble of simulations of extraction rates and permeability field realizations. 

% Why is this problem important
We demonstrate how the extraction rate can be controlled so there is a high probability that pressure at a critical location is below a given threshold. For this, we simulate an ensemble of porous media realizations and treat the numerical solutions of Darcy's equation as noisy observations of the actual solution that is approximated by a Gaussian Process Regression (GPR) surrogate model \cite{williams.GP_ML} for the pressure at the critical location. 
Tens of thousands of simulations can be required to characterize the distribution of solutions for the extraction rate and permeability field inputs. Even though a single computer simulation might only require a few minutes of computer time, an ensemble of thousands of simulations can take hours or days. 
Because practitioners often need rapid online guidance for real-time control of the injection-extraction process, we use a database of past simulations to create a surrogate model that can provide this guidance in a few seconds. 

%We create a database of simulations with different extraction rates for the ensemble of simulations characterizing the flow in the random permeability field. 

Machine learning and GPR models have proven helpful in building a surrogate between the extraction rate and  permeability field realization inputs, and the pressure at the critical location outputs. Fitting a standard GPR surrogate typically costs $\mathcal{O}(n^3)$ where $n$ is the number of data points. By strategically matching sampling locations and covariance kernels, we reduced the cost to $\mathcal{O}(n \log n)$.

Unlike most machine learning models, the GPR model provides a probability distribution over a broad class of possible solutions to provide a sound basis for quantifying the uncertainty in the predictions. For example, while neural networks may provide a prediction of equal quality, a Gaussian process provides confidence levels associated with the predictions based on similarity to already seen data.  

\begin{figure}[tb]
    \centering
    \includegraphics[width=.8\textwidth]{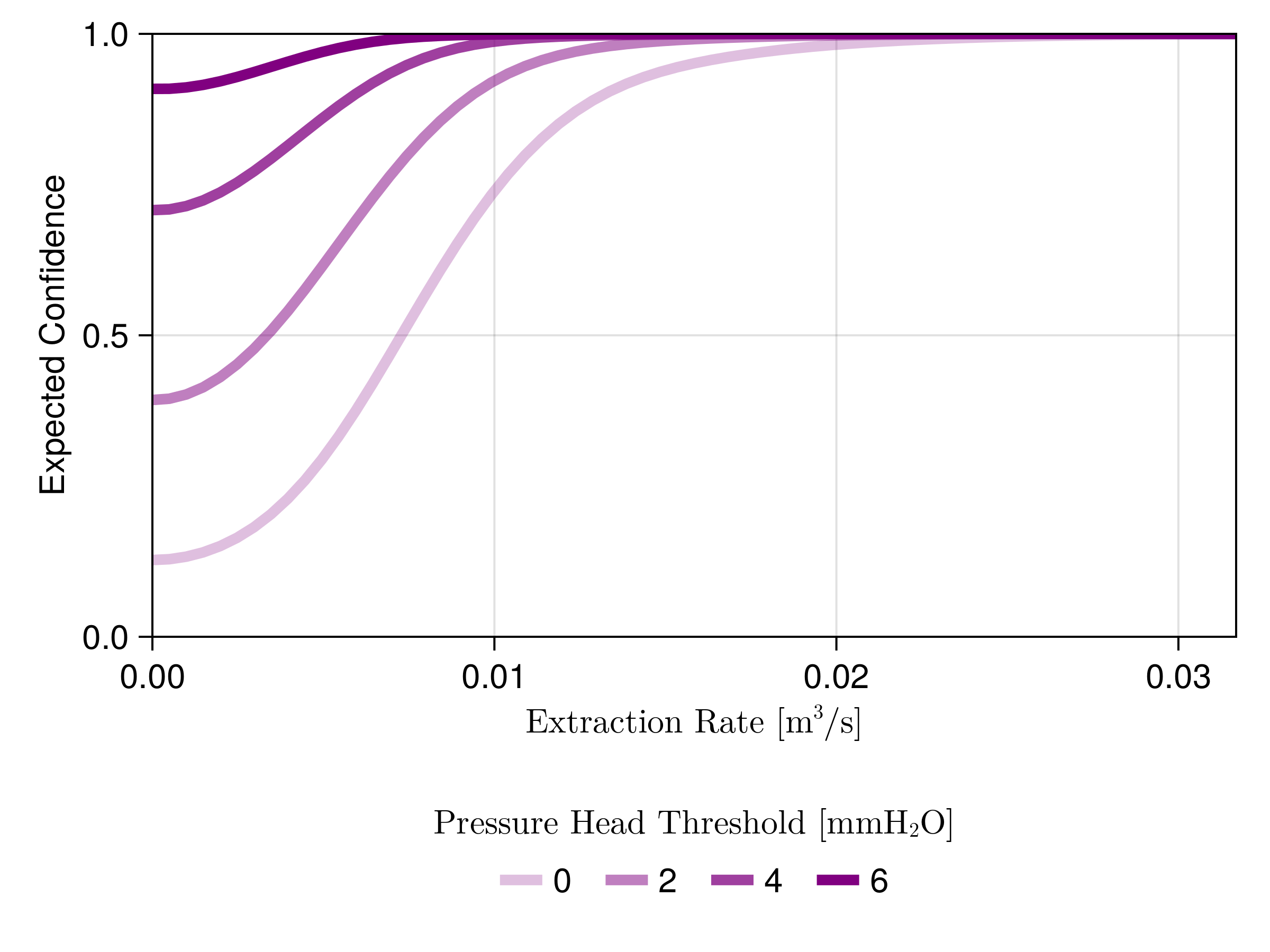}
    \caption{The expected confidence in maintaining pressure below a threshold as a function of extraction rate [m$^3$/s]. Figure \ref{fig:confidences_heatmap} extends this plot to a continuum of thresholds.}
    \label{fig:confidences_lineplot}
\end{figure}

Figure \ref{fig:confidences_lineplot} shows the GPR model prediction for the expected confidence that the pressure at the critical location will be below a given threshold. For example, suppose the threshold pressure is 2\,mmH$_2$O (millimeters of water). In this case, the extraction rate must be at least 0.01\,m$^3$/s (cubic meters per second) to have 90\% confidence that the pressure at the critical location will be below the threshold. Notice the confidence computed using the GPR surrogate increases in both the extraction rate and threshold, which matches the physics of the simulation.

%The GPR reduces the cost of fitting and evaluating the surrogate and absorbing the approximation error into the surrogate uncertainty. 

\begin{figure}[tb]
    \centering
    \includegraphics[width=.8\textwidth]{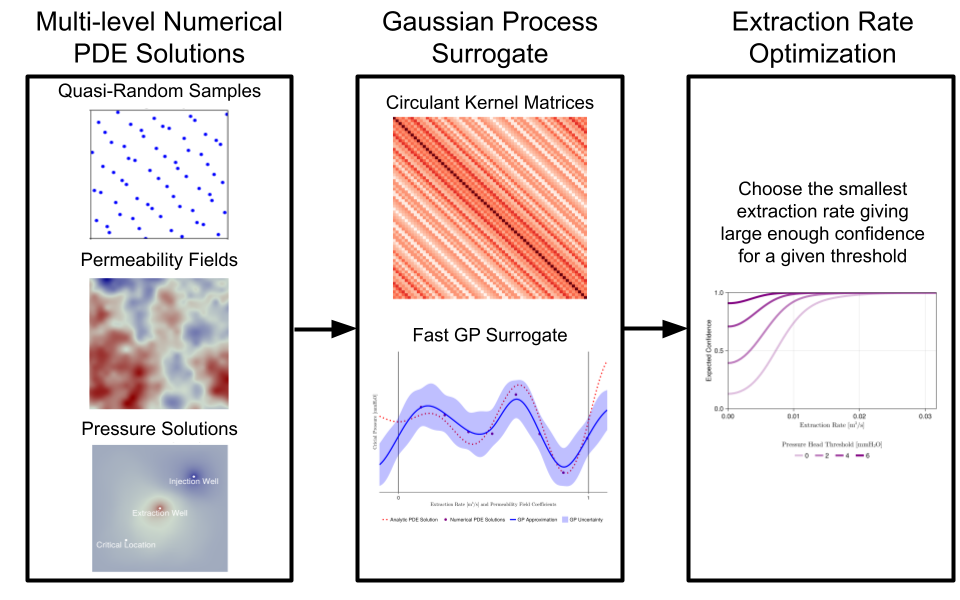}
    \caption{Workflow diagram visualizing the three stages of our method. First, the possible extraction rate and permeability field realizations are sampled with a low-discrepancy quasi-random uniform distribution. 
    Each pair is input to the numerical partial differential equation solver, which returns an approximation for the pressure at a critical location. Next, a GPR model is optimized to the data relating extraction rate and permeability to pressure at the critical location. The optimization for $n$ samples is done at $\mathcal{O}(n \log n)$ cost  by exploiting structure in the Gram kernel matrix induced by using quasi-random samples and matching kernels. The trained GPR model can be quickly evaluated to identify the lowest extraction rate, which rarely overpressurizes a critical location. %\AlekseiNote{Edit at \url{https://docs.google.com/drawings/d/1Ld3FeGcXTMeDJIHv66nASk1wpDA94knPGwOEV8waPRQ/edit?usp=sharing}}
    }
    \label{fig:subsurface_flow_workflow}
\end{figure}

Figure \ref{fig:subsurface_flow_workflow} shows the workflow of our approach:
\begin{enumerate}
\item We sample the feasibility space by generating a uniform quasi-random (low-discrepancy) sample for the extraction rates and permeability fields. 
\item We numerically solve Darcy's equation for each extraction-permeability pairing using the DPFEHM software package \cite{DPFEHM.jl}. The simulation errors depend on the fidelity of the permeability field and the fidelity of the finite volume numerical solver. 
\item With numerical solutions in hand, the observations are used to fit a fast GPR model. 
\end{enumerate}
The fitted GPR surrogate can be used to select an optimal extraction rate in real time for any pressure threshold. Our proposed workflow is general, and readily generalizes from the example we provide in this paper to other other PDEs with random coefficients.

%The workflow in this paper is presented for real-time pressure modeling at the critical location subject to a user-selected extraction rate and random permeability field. 
%However, the workflow broadly applies to problems with random permeability fields other than Darcy's equation. 

Section \ref{sec:methods} introduces the modeling equations and notation for the problem formulation and the existing methods we build upon later in the paper. This includes details on numerical solutions of Darcy's equation and an overview of GPR modeling. Our novel contributions are detailed in Section \ref{sec:novel_contributions}, where we discuss a method for calibrating surrogate error to the numerical PDE solution error. Also discussed are details on fitting a fast Gaussian process at $\mathcal{O}(n \log n)$ cost. Section \ref{sec:results} discusses implementation specifics and exemplifies the use of the trained GPR model for real-time pressure management. Finally, Section \ref{sec:conclusions} ends with a brief conclusion and discussion of future work. 

\section{Methods} \label{sec:methods}

This section describes the problem and the model equations of interest. We introduce the two-point flux finite volume method used to solve Darcy's equation and give an overview of the probabilistic GPR surrogate. 

\subsection{Problem Formulation} \label{subsec:problem_formulation}

%\subsection{Physics Model}

Consider a pressure management problem of a single-phase fluid in a heterogeneous permeability field. Darcy's partial differential equation can model the pressure throughout the subsurface 
\begin{equation}
    \nabla  \cdot (G(x)  \cdot \nabla H(x)) = f(x),
    \label{eq:darcy}
\end{equation}
when the subsurface permeability field is known. Darcy's equation describes the pressure head $H(x)$ in the subsurface over a domain $D \subset \mathbb{R}^2$ with permeability field $G(x)$ and external forcing function $f(x)$. The steady-state Darcy equation \ref{eq:darcy} allows us to evaluate the long-term impact of the injection and extraction on the pressure head. For pressure management, the forcing function $f$ is composed of an
injection rate $w \geq 0$ at $x_\text{injection}$ and an extraction rate
$-r \leq 0$ at $x_\text{extraction}$. Following 
\cite{pachalieva.pressure_management_PIML_CNN_GP}, we write 
\begin{equation}
    f(x;r) := \begin{cases} 
        w, & x = x_\text{injection} \\ 
        -r, & x = x_\text{extraction} \\
        0, & x \in D \backslash \{x_\text{injection},x_\text{extraction}\}.
    \end{cases}
    \label{eq:flow_rate}
\end{equation}
Throughout this paper, we assume that the \emph{injection rate} $w$ is fixed and focus on optimizing the \emph{extraction rate} $r$ to achieve a desired objective.  

%\subsection{Modelling the Heterogeneous Permeability Field}

The details of the permeability field $G$ are rarely known in practice. Instead, one is often given statistical properties of that field, such as the mean permeability and spatial correlations of variations of the permeability around its mean. Given these descriptors, it is convenient to model the permeability $G$ as a \emph{random Gaussian field}. Solving Darcy's equation (\ref{eq:darcy}) 
using a stochastic permeability field $G$ induces randomness in the pressure $H$.

%\subsection{Pressure at the Critical Location}

\begin{figure}[tb]
    \centering
    \includegraphics[width=.8\textwidth]{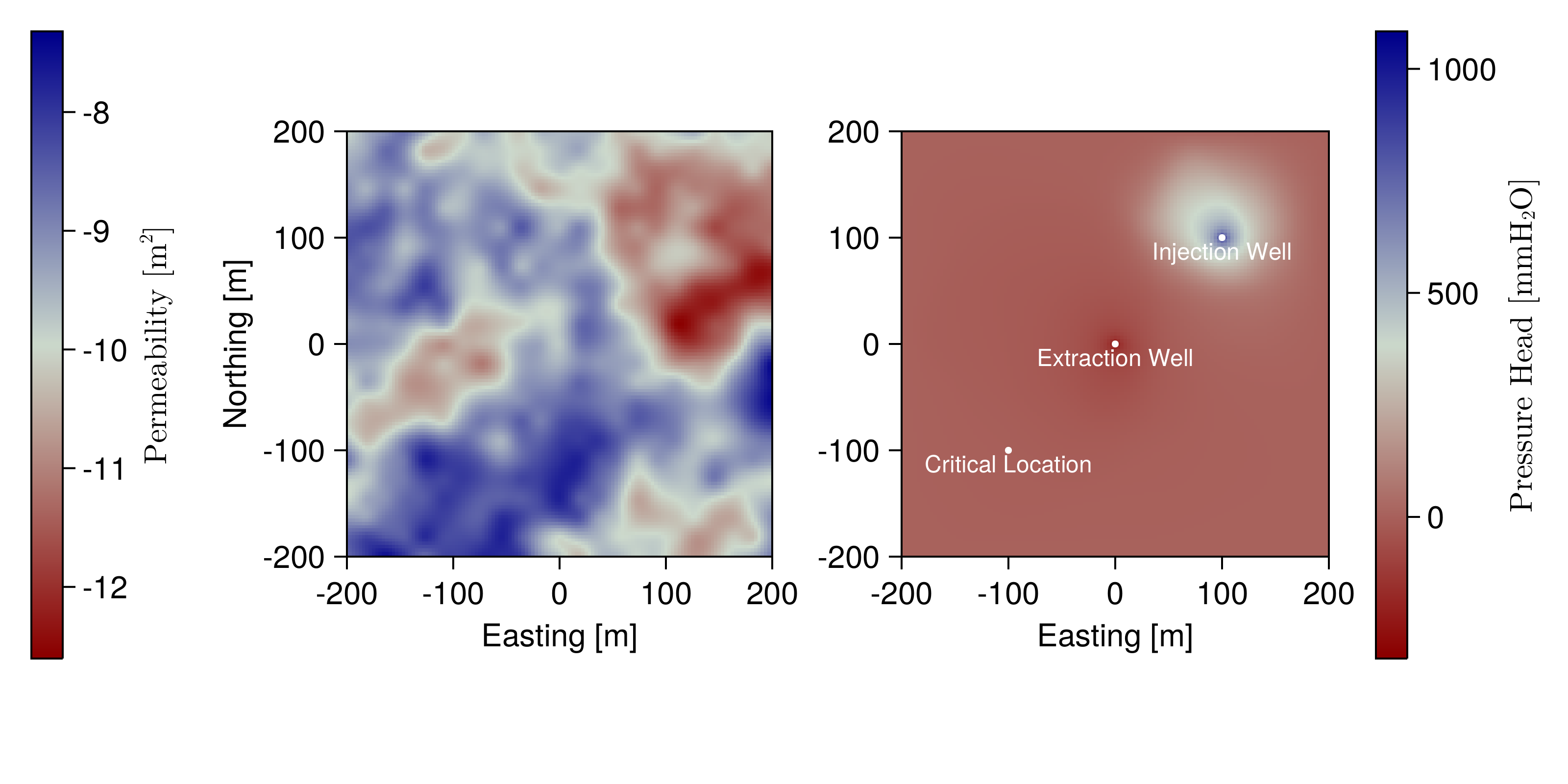}
    \caption{The left plot shows a realization of the permeability field $G(x)$ [m$^2$] and the right plot shows the resulting pressure $H(x; r,G)$ for some extraction rate $r$ [m$^3$/s]. The fixed injection, extraction, and critical locations in our two-dimensional setup are also shown.}
    \label{fig:numerical_pde.birdseye}
\end{figure}

Our goal is to quickly estimate the probability that the pressure at the \emph{critical location} $x_\text{critical}$ remains below a desired \emph{threshold} $\bar{h}$ as a function of the extraction rate $r$ to enable practitioners to implement control policies to manage pressure at critical locations in real time. 
Let $H^c(r,G) := H(x_\text{critical};r,G)$ denote the pressure at the critical location,  which is a function of the extraction rate, $r$, and Gaussian process, $G$. 
For a fixed upper bound $\bar h$, we seek to evaluate the probability
\begin{equation}
    c(r) := P_G(H^c(r,G) \leq \bar{h}),
    \label{eq:confidence}
\end{equation}
where the probability $P_G$ is taken over the distribution of the Gaussian permeability field $G$. Figure \ref{fig:numerical_pde.birdseye} illustrates the described setup. 

Although the confidence, $c(r)$, cannot be computed explicitly, it can be approximated for each fixed extraction rate $r$ by numerically solving for critical pressure $H^c(r,G)$ in Darcy's equation \eqref{eq:darcy} for many realizations of $G$. Unfortunately, this method is biased as the numerical critical pressure only approximated the critical pressure $H^c(r,G)$.  Also, the associated cost of solving the PDE multiple times is impractical for practitioners desiring fast online inference. 

%\subsection{GPR surrogate and Induced Confidence}

Our approach provides rapid solutions and error estimates for the confidence $c(r)$ by building a surrogate model for the critical pressure $H^c(r,G)$. This statistical approach treats the numerically computed critical pressures as noisy observations of the analytic critical pressures. GPR is a natural and efficient approach for this framework and can provide immediate online estimates for $c(r)$ as a function of the extraction rate. 

Given $n$ numerical critical pressure observations, the GPR surrogate $H^c_n(r,G)$ estimates the critical pressure $H^c(r,G)$. We plug this estimate into \eqref{eq:confidence} and get the \emph{conditional confidence} 
\begin{equation}
    \hat{C}_n(r) := P_G(H^c_n(r,G) \leq \bar{h} | H_n^c).
    \label{eq:conditional_confidence}
\end{equation}
The \emph{expected conditional confidence} is a natural estimate for $c(r)$ denoted by 
\begin{equation}
    c_n(r) := \mathbb{E}_{H_n^c} \left[\hat{C}_n(r)\right] = P_{(G,H_n^c)}(H^c_n(r,G) \leq \bar{h}).
    \label{eq:expected_conditional_confidence}
\end{equation}
We approximate the unknown analytic solution $c(r)$ by the computationally tractable $c_n(r)$, which only uses the surrogate model. After interchanging expectations, the above equation can be efficiently computed with (Quasi-)Monte Carlo.

\subsection{Numerical Solution of Darcy's Equation}

To solve Darcy's equation \eqref{eq:darcy}, we apply a standard two-point flux finite volume method in the domain $D$ on a discrete mesh. A truncated Karhunen-Loève expansion represents the Gaussian permeability field $G$ over $D$. This enables us to draw samples of $G$, which can be evaluated at a mesh grid of any fidelity. 

We say the physical domain has discretization dimension $d$ when the finite volume mesh has $d+1$ mesh points in each dimension of $D$. The choice of $d=2^m$ creates nested mesh grids. While there is no restriction on the mesh grids with an equal number of points in each dimension, this reduces the number of parameters we must consider when approximating the numerical error later in this section. 

\begin{figure}[tb]
    \centering
    \includegraphics[width=0.8\textwidth]{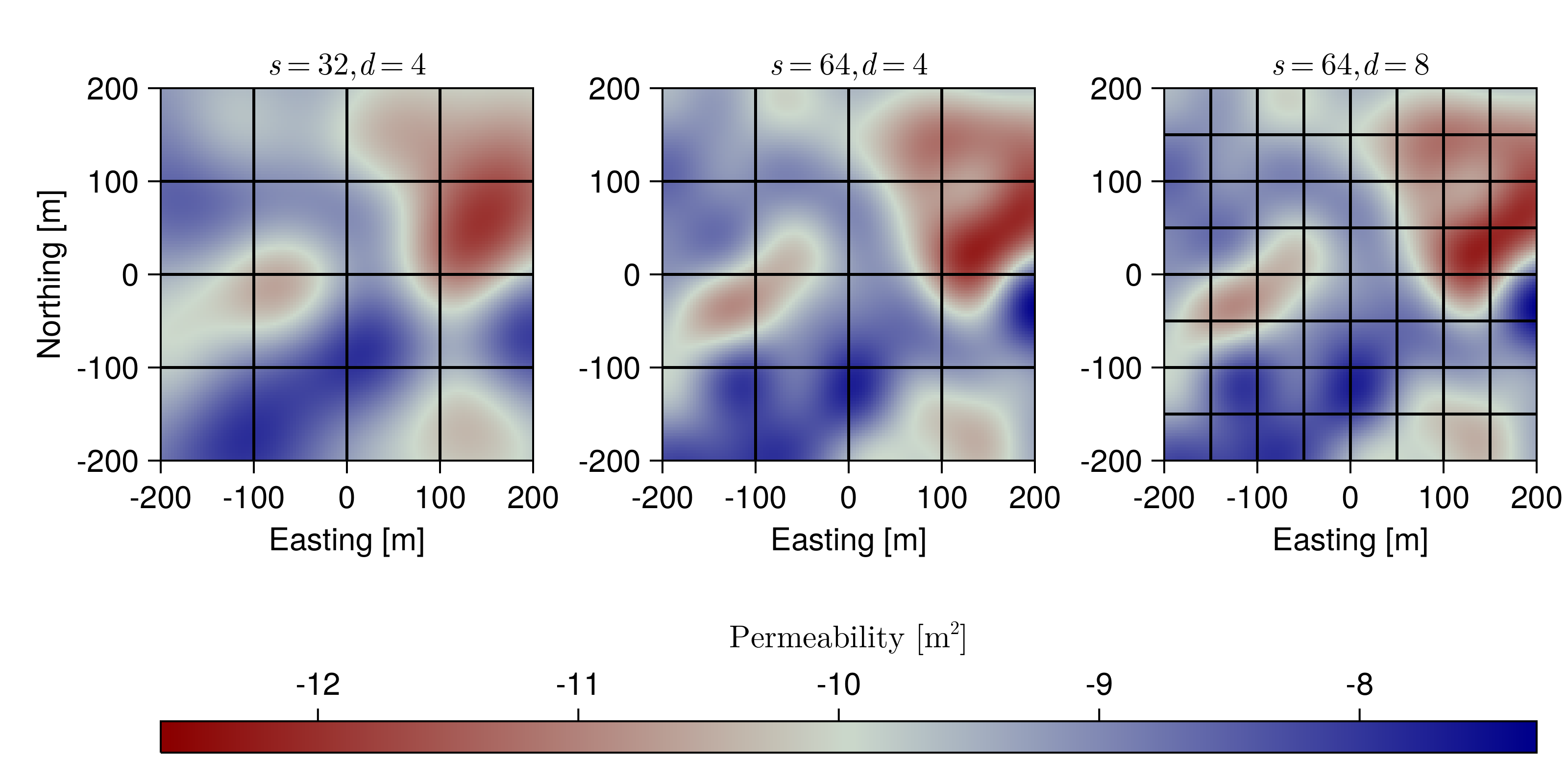}
    \caption{The same realization of the permeability field as in Figure \ref{fig:numerical_pde.birdseye} but for various choices of permeability discretization dimension $s$ and domain discretization dimension $d$. In the left plot, a small $s$ and $d$ are chosen, corresponding to a lack of small-scale changes in the permeability realization and a coarse mesh grid over the domain, respectively. Moving from the left to center plot maintains the same mesh grid while increasing $s$ to yield more small-scale changes in the realization. Moving from the center to the right plot keeps the same realization while increasing $d$ to yield a finer mesh over $D$.}
    \label{fig:numerical_pde.full}
\end{figure}

The Karhunen-Loève expansion \cite{karhunen.kl_expansion} %huang.convergence_kl,ghanem.stochastic_finite_elements_kl,shinozuka.simulation_stochastic_processes,grigoriu.spectral_represetation_simulation
of the permeability field may be used to find a good finite-dimensional approximation of $G$. Specifically, any Gaussian random field can be represented as 
\begin{equation}
    G(x) = \sum_{j=1}^\infty \sqrt{\lambda_j} \varphi_j(x) Z_j
    \label{eq:KL_expansion}
\end{equation}
where $\varphi_j$ are deterministic and orthonormal and $Z_1, Z_2, \dots$ are independent standard Gaussian random variables. Ordering $\lambda_1 \geq \lambda_2 \geq \lambda_3 \geq \dots$, we approximate $G$ by 
\begin{equation}
    G^s(x) := \sum_{j=1}^s \sqrt{\lambda_j} \varphi_j(x) Z_j 
    \label{eq:KL_expansion_approx}
\end{equation}
which optimally compacts the variance into earlier terms. Figure \ref{fig:numerical_pde.full} shows different pairs of $s$ and $d$ for a common realization of $Z_1,Z_2,\dots$. Notice the greater detail in $G^s$ as $s$ increases and the finer mesh over $D$ as $d$ increases. 

We let $H^c_{s,d}(r,\boldsymbol{Z}^{s})$ denote the \emph{numerical critical pressure} computed by solving the PDE \eqref{eq:darcy} with domain discretization dimension $d$ and permeability discretization dimension $s$. Here $\boldsymbol{Z}^{s} = (Z_1,\dots,Z_s)$ is a vector of independent standard Gaussians, which uniquely determine the approximate permeability field $G^s$. In our implementation, we use the \texttt{GaussianRandomFields.jl} package \cite{GaussianRandomFields.jl} to simulate permeability fields $G^s$ and solve the PDE numerically with the \texttt{DPFEHM.jl} \cite{DPFEHM.jl}. 

\subsection{A Probabilistic GPR Surrogate} \label{sec:GPs}

We model the relationship between the inputs of extraction rate $r$ and permeability field $G$ and the output critical pressure $H^c(r,G)$. The model is built on observed numerical critical pressures $Y^n := \{H^c_{s,d}(r_i,\boldsymbol{Z}_i^s)\}_{i=1}^n$ at strategically chosen sampling locations $(r_i,G_i^s)_{i=1}^n$. Our GPR surrogate views $Y^n$ as noisy observations of $H^c$ with 
\begin{equation}
    H_{s,d}^c(r,\boldsymbol{Z}^s) = H^c(r,G) + \varepsilon.
    \label{eq:gp_model}
\end{equation}
The \emph{noise} $\varepsilon$ is assumed to be a zero mean Gaussian random variable with variance $\zeta_{s,d}$. The noise encodes the discretization error and is assumed to be independent of the sampling location $(\boldsymbol{Z}^s,G)$ but dependent on the discretizations.

GPR assumes $H^c$ is a Gaussian process, and therefore, the conditional distribution of $H^c$ given $Y^n$ is also a Gaussian process \cite{williams.GP_ML}. We use this conditional, or posterior, distribution on $H^c$ as an error-aware surrogate. The conditional mean and covariance functions determining the posterior Gaussian process are
\begin{align}
    m_n(t) &:= \mathbb{E}\left[H^c(t) | Y^n\right] \label{eq:post_mean} \quad \text{and} \\
    k_n(t,t') &:= \mathrm{Cov}\left[H^c(t),H^c(t') | Y^n \right] \label{eq:post_cov}
\end{align}
respectively where $t := (r,\boldsymbol{Z}^s)$ are the GPR inputs. The conditional variance is written as 
\begin{equation}
    \sigma_n^2(t) := \mathrm{Var}\left[H^c(t) | Y^n \right] = k_n(t,t). 
    \label{eq:post_var}
\end{equation}
Building upon the above notation, we denote the posterior Gaussian process by 
\begin{equation}
    H_n^c := H^c | Y_n \sim \mathrm{GP}(m_n,k_n).
    \label{eq:post_gp}
\end{equation}

\begin{figure}[tb]
    \centering
    \includegraphics[width=.8\textwidth]{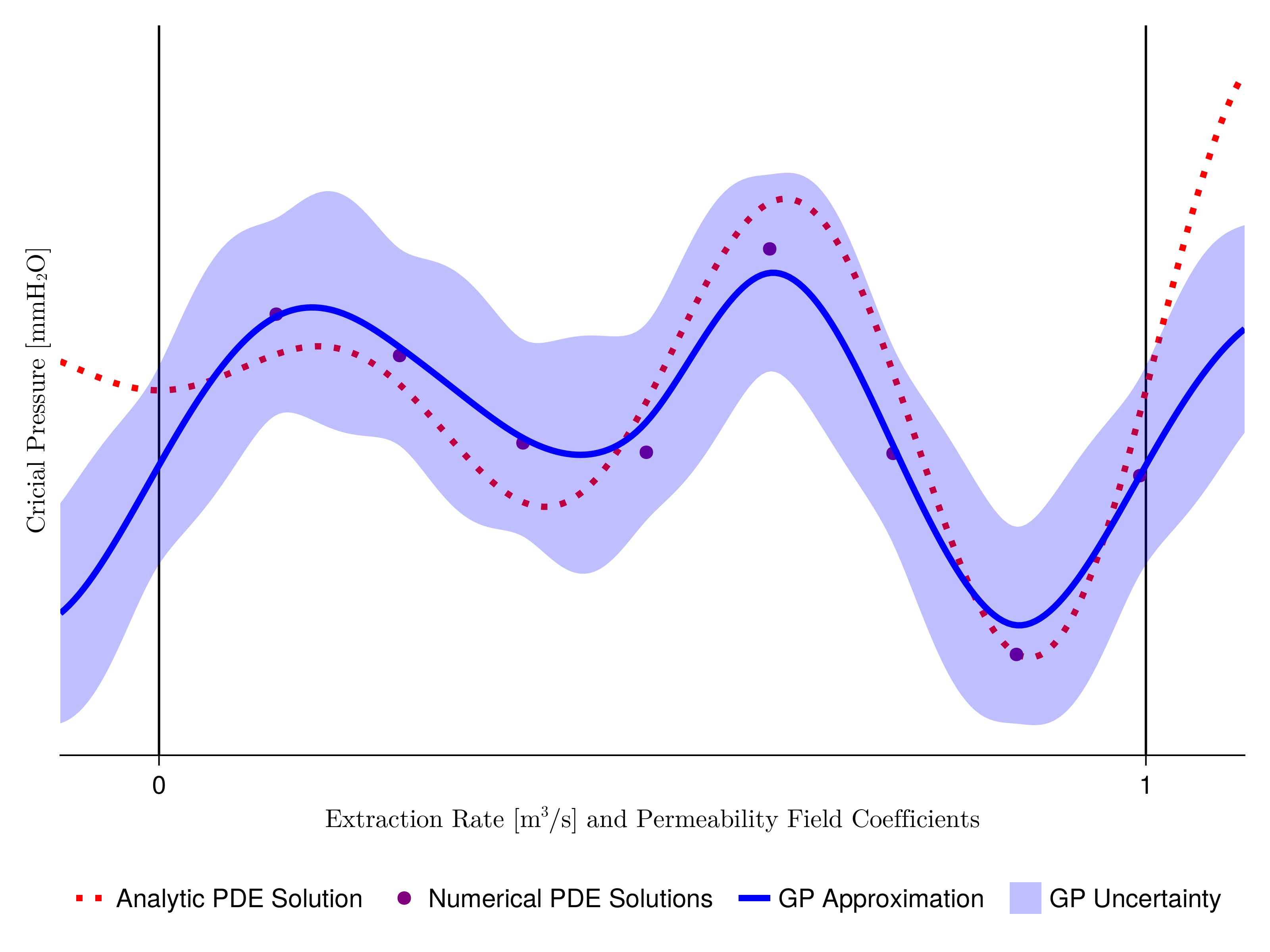}
    \caption{Cartoon Gaussian process model. Suppose we are interested in recovering the analytic PDE solution $H^c$ from numerical PDE solutions, which we treat as noisy observations of $H^c$. The posterior Gaussian process (GP) for $H^c$ is visualized through its posterior mean approximation and 99\% confidence interval uncertainty at each point. While the cartoon shows a one-dimensional input, our actual model takes a  $1+s$ dimensional input: $1$ for the extraction rate and $s$ for the random coefficients in the Karhunen-Loève expansion of the permeability field.}
    \label{fig:noisy_lattice_qgp.4}
\end{figure}

Figure \ref{fig:noisy_lattice_qgp.4} illustrates an example posterior Gaussian process. While the figure assumes $t$ is one-dimensional, which is impossible for our problem, GPR extends naturally to arbitrarily large dimensions. We emphasize that the posterior Gaussian process is a surrogate for the critical pressure $H^c$, not the numerical critical pressure $H^c_{s,d}$ whose evaluations are used for fitting. Our GPR surrogate provides a point estimate for the critical pressure and the distribution on $H^c$.

\section{Theory} \label{sec:novel_contributions}

This section describes in detail the novel contributions of this work. First, we derive an upper bound on the root mean squared error between the numerical PDE solution and the true PDE solution. This bound is used as a conservative initial guess of the GPR noise variance before hyperparameter optimization. Second, we discuss how the fitting of the GPR model to $n$ data points can be accelerated from $\mathcal{O}(n^3)$ to $\mathcal{O}(n \log n)$ using an intelligent design of experiments and matching GPR kernel.

\subsection{Approximate Upper Bound on Numerical Error}

We approximate an upper bound on the Root Mean Squared Error (RMSE) between the target and numerical critical pressures. This upper bound is used as a starting point to calibrate the noise variance in the GPR model. 

For permeability discretization dimension $s$ and domain discretization dimension $d$, the RMSE is 
\begin{equation}
    \begin{aligned}
        \mathrm{RMSE}_{s,d} &:= \left\lVert H^c(R,G)-H^c_{s,d}(R,\boldsymbol{Z}^{s}) \right\rVert \\
        &= \sqrt{\mathbb{E}_{\left(R,\boldsymbol{Z}^{s}\right)} \left[H^c(R,G)-H^c_{s,d}(R,\boldsymbol{Z}^{s})\right]^2}~.
    \end{aligned}
    \label{eq:post_var_N_as_expected_val}
\end{equation}
Here the extraction rate $R \sim \mathcal{U}[0,w]$ is assumed to be independent of $\boldsymbol{Z}^{s}$.

Following ideas from Multi-level Monte Carlo \cite{giles.multilevel_mc_path_simulation,robbe.multi_index_qmc_lognormal_diffusion}, we choose strictly increasing sequence $(s_j)_{j \geq 0}$ and $(d_j)_{j \geq 0}$ and rewrite \eqref{eq:post_var_N_as_expected_val} as the telescoping sum 
\begin{equation}
    \mathrm{RMSE}_{s_N,d_N} = \left\lVert \sum_{j=N+1}^\infty \left[\Delta_{s_j}(R,\boldsymbol{Z}^{s_j}) + \Delta_{d_j}(R,\boldsymbol{Z}^{s_j}) \right]\right\rVert
    \label{eq:telescoping_post_var_N}
\end{equation}   
where 
\begin{align*}
    \Delta_{s_{j+1}}(R,\boldsymbol{Z}^{s_{j+1}}) &= H^c_{s_{j+1},d_j}(R,\boldsymbol{Z}^{s_{j+1}}) - H^c_{s_j,d_j}(R,\boldsymbol{Z}^{s_j}) \quad \text{and} \\
    \Delta_{d_{j+1}}(R,\boldsymbol{Z}^{s_{j+1}}) &= H^c_{s_{j+1},d_{j+1}}(R,\boldsymbol{Z}^{s_{j+1}}) - H^c_{s_{j+1},d_j}(R,\boldsymbol{Z}^{s_{j+1}}).
\end{align*}
Let us assume that 
\begin{equation}
    \lVert \Delta_{s_j}(R,\boldsymbol{Z}^{s_j}) \rVert = 2^{b_s} s_j^{a_s} \quad\text{and}\quad  \lVert \Delta_{d_j}(R,\boldsymbol{Z}^{s_j}) \rVert = 2^{b_d}d_j^{a_d}.
    \label{eq:logloglinear_delta_dims}
\end{equation}
The parameters $(a_s,b_s)$ and $(a_d,b_d)$ will be fit using linear regression in the log-log domain. Let $s_j = v_s 2^j$ and $d_j = v_d 2^j$ where $v_s$ and $v_d$ are initial values for discretization dimension $s$ and $d$ respectively which are chosen by the user. Applying the triangle inequality to \eqref{eq:telescoping_post_var_N} gives 
\begin{equation}
    \begin{aligned}
        \mathrm{RMSE}_{s_N,d_N} &\leq \sum_{j=N+1}^\infty \left[2^{b_s}v_s^{a_s} \left(2^{a_s}\right)^j + 2^{b_d} v_d^{a_d} \left(v_d^{a_d}\right)^j\right] \\
        &= 2^{b_s} v_s^{a_s} \frac{2^{(N+1) a_s}}{1-2^{a_s}} + 2^{b_d}v_d^{a_d} \frac{2^{(N+1) a_d}}{1-2^{a_d}} \\
        &=: \overline{\mathrm{RMSE}}_{s_N,d_N}
    \end{aligned}
    \label{eq:bar_rmse_upper_bound}
\end{equation}
using the expression for the sum of a geometric series.

\begin{figure}[tb]
    \centering
    \includegraphics[width=.8\textwidth]{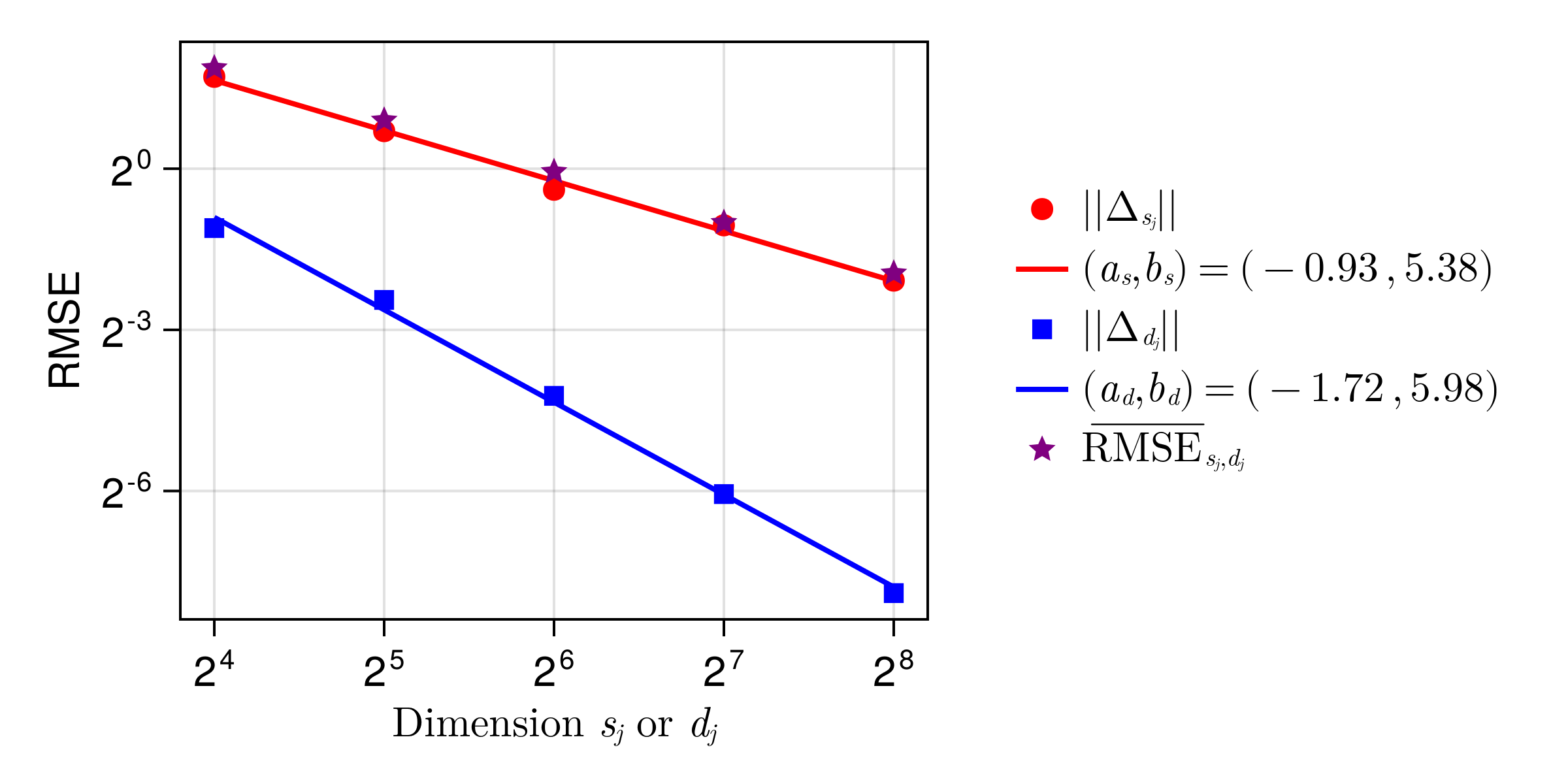}
    \caption{The Karhunen-Loève expansion of the permeability field is approximated by the sum of the first $s_j$ terms. The physical domain $D$ is discretized into a grid with mesh width $1/d_j$ in each dimension. The RMSE of the difference between numerical critical pressures with discretizations $(s_j,d_j/2)$ and $(s_j/2,d_j/2)$ is scatter plotted as $\lVert \Delta_{s_j} \rVert$. The RMSE of the difference between numerical critical pressures with discretizations $(s_j,d_j)$ and $(s_j,d_j/2)$ is scatter plotted as $\lVert \Delta_{d_j} \rVert$.  Simple regression models are fit to these two scatter trends where $a$ and $b$ are the slope and intercept, respectively, of the plotted lines. These models are extrapolated through an infinite telescoping sum to derive the approximate upper bound on the RMSE of the difference between the numerical critical pressure with discretization $(s_j,d_j)$ and the target critical pressure.}
    \label{fig:convergence}
\end{figure}

Figure \ref{fig:convergence} illustrates the above idea. At every dimension pair $(s_j,d_j)$ we solve the PDE at the same $(R_i,\boldsymbol{Z}_i^{s_j})_{i=1}^m$ points to get $\{H^c_{s_j,d_j}(R_i,\boldsymbol{Z}_i^{s_j})\}_{i=1}^m$. For $j=1,\dots,N$ we make the approximations 
\begin{equation}
    \begin{aligned}
        \lVert \Delta_{s_j}(R,\boldsymbol{Z}^{s_j}) \rVert &\approx \sqrt{\frac{1}{m} \sum_{i=1}^m \Delta^2_{s_j}(R_i,\boldsymbol{Z}^{s_j})}, \\
        \lVert \Delta_{d_j}(R,\boldsymbol{Z}^{s_j}) \rVert &\approx \sqrt{\frac{1}{m} \sum_{i=1}^m \Delta^2_{d_j}(R_i,\boldsymbol{Z}^{s_j})} 
    \end{aligned}
    \label{eq:RMSEs_approxes}
\end{equation}
corresponding to the plotted red dots and blue squares, respectively. The slope intercept pairings $(a_s,b_s)$ and $(a_d,b_d)$ from \eqref{eq:logloglinear_delta_dims} are fit to the values in \eqref{eq:RMSEs_approxes} with lines in the respective colors. The upper bounds $\overline{\mathrm{RMSE}}_{s_j,d_j}$ from \eqref{eq:bar_rmse_upper_bound} are visualized by the purple stars. Notice that the model will find $\overline{\mathrm{RMSE}}_{s_j,d_j}$ for any dimension pair $(s_j,d_j)$ we choose. As $d_j$ increases, the mesh size shrinks, and the PDE becomes more expensive to solve numerically. As $s_j$ increases, the input dimension to the model grows, and more PDE solves are required to build an accurate model. Notice that the RMSE is dominated by the error in the permeability field discretization rather than the error in the domain discretization.  

\subsection{Fast GPR}

Gaussian processes regression models are defined with a positive-definite covariance kernel $k$, which assumes $\mathrm{Cov}[H^c(t),H^c(t')] = k(t,t')$. The posterior mean $m_n$ and posterior covariance $k_n$ given the data $Y^n$ requires solving the linear system $K a = b$ for $a \in \mathbb{R}^n$ given $b \in \mathbb{R}^n$ and $K = (k(t_i,t_j))_{i,j=1}^n$, a $n \times n$ matrix of the kernel $k$ evaluated at pairs of sampling locations $T := (t_i)_{i=1}^n$. The cost of solving this system is $\mathcal{O}(n^3)$ in the general case where $K$ is dense and unstructured. This computational cost limits the sample size used to build the surrogate. 

For some structured $K$, the linear system can be solved in $\mathcal{O}(n \log n)$. For example, when $K$ is circulant- or block-Toeplitz \cite{gray2006toeplitz}, the linear system can be solved using fast Fourier transforms. We can induce such structure in $K$ by strategically choosing the sampling locations $T$ and matching covariance kernel $k$. Two available flavors are: 
\begin{itemize}
    \item Lattice sequence $T$ and periodic shift invariant $k$ produce circulant $K$ \cite{rathinavel.bayesian_QMC_lattice}. 
    \item Digital sequence $T$ and digitally shift invariant $k$ produce block Toeplitz $K$ \cite{rathinavel.bayesian_QMC_sobol}. 
\end{itemize}
A unifying overview of these methods can be found in \cite{rathinavel.bayesian_QMC_thesis}. The details for kernel interpolation in reproducing kernel Hilbert space are described in \cite{kaarnioja.kernel_interpolants_lattice_rkhs,kaarnioja.kernel_interpolants_lattice_rkhs_serendipitous,kuo.kernel_interpolants_lattice_+rkhs_search_criteria}. 

\begin{figure}[tb]
    \centering
    \includegraphics[width=.8\textwidth]{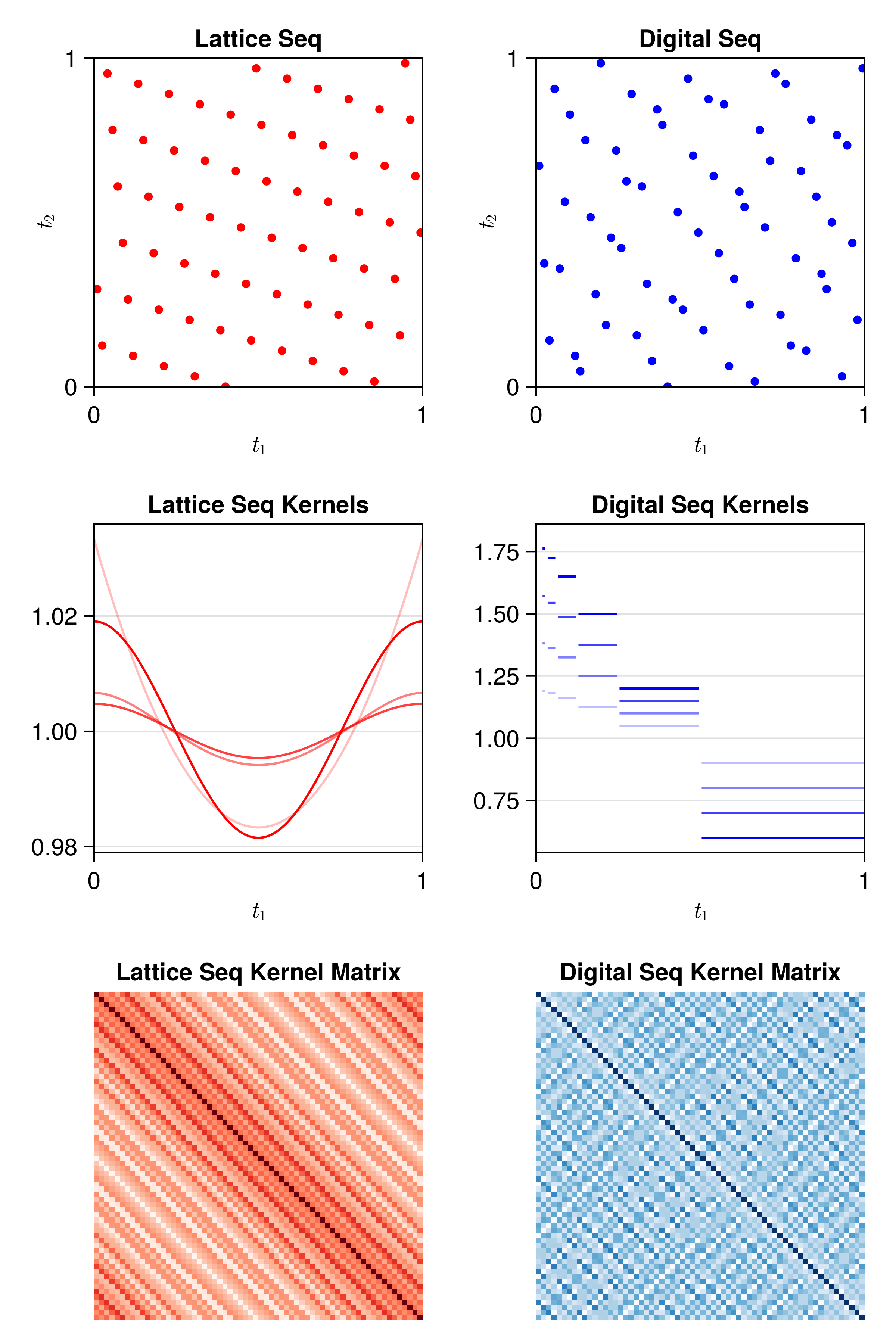}
    \caption{The first column shows the first $n$ points of a lattice sequence and some matching periodic kernels that yield circulant kernel matrices. The second column shows the first $n$ points of a digital sequence and some matching step function kernels that yield block Toeplitz kernel matrices. Features of the kernel functions are induced in the prediction function. For instance, using the lattice sequence and matching kernels will yield a periodic prediction, while using digital sequences and matching kernels will yield a discontinuous step-function prediction.}
    \label{fig:qgp}
\end{figure}

Figure \ref{fig:qgp} plots lattice and digital sequences $T$ alongside matching kernels $k$ and the induced kernel matrices $K$. Notice that lattice and digital sequences lie in the unit cube and have low discrepancy with the standard uniform distribution \cite[Chapters 15 and 16]{owen.MC_book}. The periodicity in the lattice sequence kernels and discontinuities in the digital sequence kernels induce the same features in the posterior mean. For example, Figure \ref{fig:noisy_lattice_qgp.4} showed a GPR surrogate with lattice sampling locations and matching kernel. 

Table \ref{table:gp_costs} compares construction and evaluation costs between the unstructured and structured Gaussian processes described in this section. The construction costs include the tuning parameters to maximize the marginal likelihood of the Gaussian process and the computing constants to be used during evaluation. Evaluation costs for both the posterior mean and covariance occur after construction. 

\begin{table}[ht]
    \centering
    \begin{tabular}{r|cccc}
        & \multicolumn{2}{c}{Construction} & \multicolumn{2}{c}{Evaluation} \\
        & Tuning  & Constants & Mean & Covariance \\
        \hline
        Unstructured GP & $\mathcal{O}(n^3)$ & $\mathcal{O}(n^3)$ & $\mathcal{O}(n)$ & $\mathcal{O}(n^2)$ \\
        Structured GP & $\mathcal{O}(n \log n)$ & $\mathcal{O}(n \log n)$ & $\mathcal{O}(n)$ & $\mathcal{O}(n \log n)$
    \end{tabular}
    \caption{Comparison of construction and evaluation costs between the unstructured and structured Gaussian processes in terms of the number of samples $n$ used to fit the Gaussian process. Construction costs include tuning parameters and computing constants for evaluating the posterior mean and covariance.}
    \label{table:gp_costs}
\end{table}

Recall that the noise variance $\zeta_{s,d}$ encodes the approximation error and acts as a regularization of the posterior Gaussian processes, which honors observations $Y^n$ less as $\zeta_{s,d}$ grows. We choose $\zeta_{s,d}$ to optimize the marginal likelihood of the Gaussian process. The upper bound $\overline{\mathrm{RMSE}}_{s_j,d_j}$ in \eqref{eq:bar_rmse_upper_bound} is used as a conservative initial noise variance $\zeta_{s,d}$ for optimizing the likelihood. We expect the optimization routine to decrease $\zeta_{s,d}$ from the conservative initial guess to a well-calibrated value. 

\section{Results} \label{sec:results}

To enable the reproducibility of our results, we describe the specifics for approximating the confidence $c(r)$ in \eqref{eq:confidence} by the expected conditional confidence $c_n(r)$ in \eqref{eq:expected_conditional_confidence}. Recall from Section \ref{sec:GPs} that lattice sequences are defined in the unit cube $[0,1]^s$. Moreover, the posterior mean of a GPR surrogate fit with lattice sampling locations, and a matching covariance kernel has a periodic posterior mean. We first transform the extraction rate $r$ and independent Gaussians $\boldsymbol{Z}^s$, defining the permeability field $G^s$ to the unit cube domain. We then periodize the critical pressure using the baker transform. Finally, we show some numerical results from applying our complete method. 

To transform our problem to the unit cube $[0,1]^s$, we first restrict the extraction rate to be no larger than the injection rate i.e. $r \in [0,w]$. Let $\Phi^{-1}$ denote the inverse distribution function of standard Gaussian random variable. For any $r_u \in [0,1]$ and $u_1,u_2,\dots \in (0,1)$ we may use \eqref{eq:KL_expansion} to write 
\begin{equation}
    M^c(r_u,u_1,u_2,\dots) := H^c\left(wr_u,\sum_{j \geq 1} \sqrt{\lambda_j} \varphi_j(x) \Phi^{-1}(u_j)\right).
\end{equation}
For $U_1,U_2,\dots \overset{\mathrm{iid}}{\sim} \mathcal{U}[0,1]$ we have 

$$P_G(H^c(r,G) \leq \bar{h}) = P_{(U_1,U_2,\dots)}(M^c(r/w,U_1,U_2,\dots) \leq \bar{h}).$$ 

To periodize $M^c(r_u,u_1,u_2,\dots)$, define the baker transform \cite[Chapter 16]{owen.MC_book} 
\begin{equation}
    b(u) = 1 - 2 \left\lvert u - \frac{1}{2}\right\rvert = \begin{cases} 2u, & 0 \leq u \leq 1/2 \\ 2(1-u) & 1/2 \leq u \leq 1 \end{cases}
\end{equation}
so that 
\begin{equation}
    \mathring{M}^c(r_u,u_1,u_2,\dots) := M^c(b(r_u),b(u_1),b(u_2),\dots).
\end{equation}
Since $b(U) \sim \mathcal{U}[0,1]$ when $U \sim \mathcal{U}[0,1]$, we have 
\begin{equation}
    \begin{aligned}
        c(r) &= P_{(U_1,U_2,\dots)}\left(M^c(r/w,U_1,U_2,\dots) \leq \bar{h}\right) \\
        &= P_{(U_1,U_2,\dots)}\left(\mathring{M}^c(r/(2w),U_1,U_2,\dots) \leq \bar{h}\right).
    \end{aligned}
    \label{eq:periodic_confidence}
\end{equation}

Let $\mathring{M}^c_n(r_u,u_1,\dots,u_s)$ denote the posterior Gaussian process for $\mathring{M}^c(r_u,u_1,u_2,\dots)$. Substituting  $\mathring{M}^c_n(r_u,u_1,\dots,u_s)$ into \eqref{eq:periodic_confidence} and taking the expectation gives 
\begin{equation}
    \begin{aligned}
        \mathring{c}_n(r) &:= P_{(\mathring{M}^c_n,U_1,\dots,U_s)}(\mathring{M}^c_n(r/(2w),U_1,\dots,U_s) \leq \bar{h}) \\
        &= \mathbb{E}_{(U_1,\dots,U_s)}\left[\Phi\left(\frac{\bar{h} - \mathring{m}_n(r,U_1,\dots,U_s)}{\mathring{\sigma}_n(r,U_1,\dots,U_s)}\right)\right].
    \end{aligned}
    \label{eq:periodic_expected_conditional_confidence}
\end{equation}
Equation \eqref{eq:periodic_confidence} motivates us approximating $c(r)$ by $\mathring{c}_n(r)$. Here the final inequality follows from Fubini's theorem \cite{fubini1907sugli} and $\mathring{m}_n$ and $\mathring{\sigma}_n$ are the posterior mean and standard deviation of $\mathring{M}_n^c$  when plugged into  \eqref{eq:post_mean} and \eqref{eq:post_var}  respectively. 

We use QMCGenerators.jl \cite{QMCGenerators.jl} to generate lattice (or digital) sequences and use QuasiGaussianProcesses.jl \cite{QuasiGaussianProcesses.jl} to fit the fast Gaussian processes. The lattice or digital sequences can also efficiently approximate \eqref{eq:periodic_expected_conditional_confidence} using Quasi-Monte Carlo \cite{niederreiter.qmc,owen.MC_book}. Specifically, we use the Quasi-Monte Carlo estimate 
\begin{equation}
    \mathring{c}_n(r) \approx \frac{1}{N} \sum_{i=1}^N \Phi\left(\frac{\bar{h} - \mathring{m}_n(r,U_{i1},\dots,U_{is})}{\mathring{\sigma}_n(r,U_{i1},\dots,U_{is})}\right)
    \label{eq:qmc_estimate_expected_conf}
\end{equation}
where $(U_{i1},\dots,U_{is})_{i=1}^N$ are low-discrepancy points e.g. the first $N$ points of a lattice or digital sequence. 

The domain of our subsurface is square with side lengths of 200\,m with the injection well, extraction well, and critical location shown in Figure \ref{fig:numerical_pde.birdseye}. We set the injection rate to 0.031688\,m$^3$/s (equivalent to 1 million metric tons per year [MMT/y]) and test extraction rates between -0.031688\,m$^3/s$ and 0.0\,m$^3$/s. We use a zero mean Gaussian permeability field with a Matérn covariance kernel having a correlation length 50\,m.

\begin{figure}[tb]
    \centering
    \includegraphics[width=0.8\textwidth]{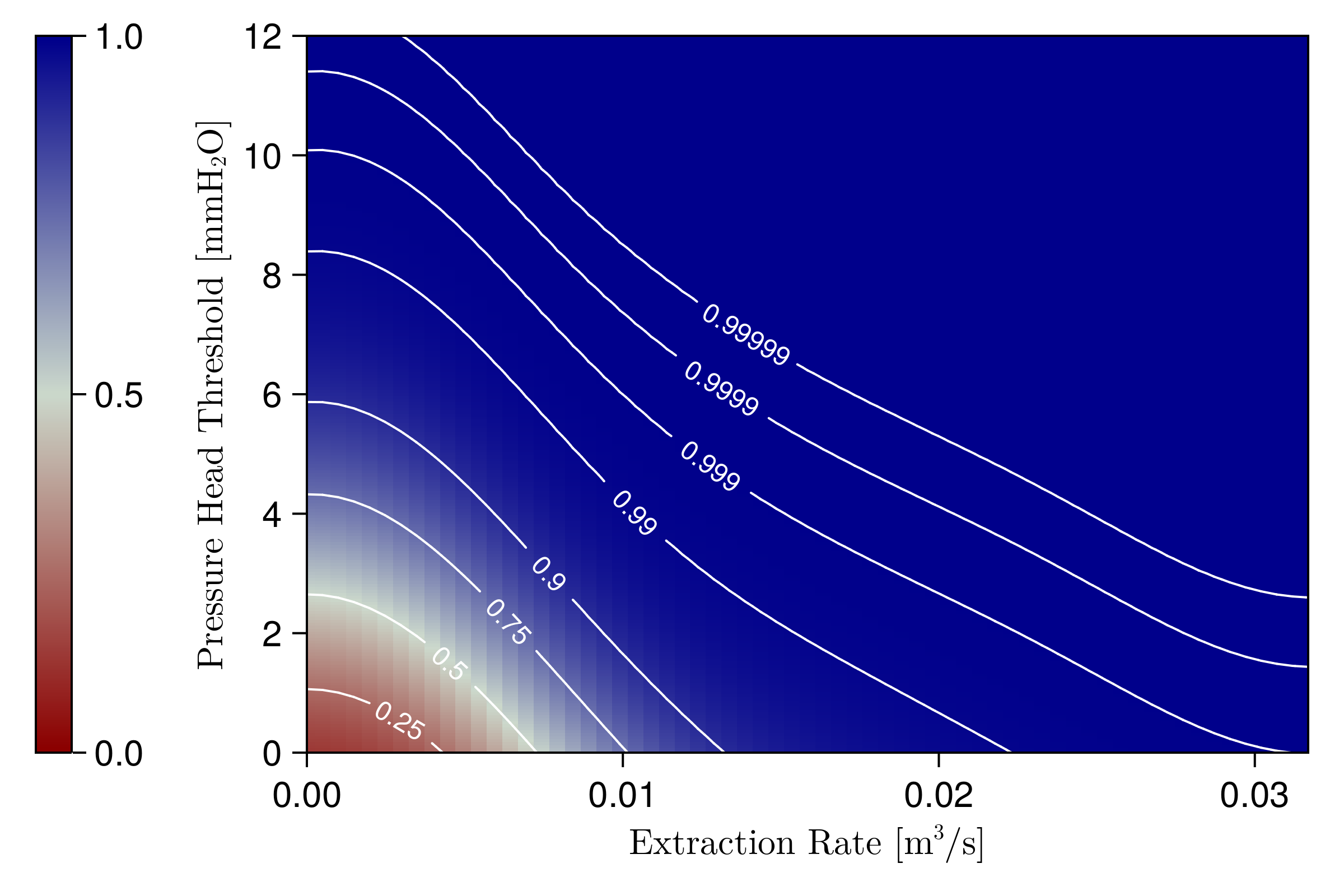}
    \caption{Approximate expected posterior confidence from \eqref{eq:qmc_estimate_expected_conf} by extraction rate $r$ and pressure threshold $\bar{h}$.}
    \label{fig:confidences_heatmap}
\end{figure}

In Figure \ref{fig:confidences_heatmap}, we plot the approximate expected conditional confidence in \eqref{eq:qmc_estimate_expected_conf} for a range of extraction rates $r$ and pressure thresholds $\bar{h}$. While the surrogate is not constrained to be monotonically increasing in both extraction rate $r$ and threshold $\bar{h}$, the expected confidence appears to have this qualitative behavior. This reassures us that our surrogate captures the physics in the model. To improve the quality of the approximation, when needed, one can: 
\begin{itemize}
    \item Fit a surrogate to higher fidelity discretizations i.e., increase $s$ and $d$.
    \item Increase $n$, the number of samples used to fit the surrogate.
    \item Increase $N$, the number of points used in the Quasi-Monte Carlo \eqref{eq:qmc_estimate_expected_conf}. 
\end{itemize}

\section{Conclusions} \label{sec:conclusions}

We fit a GPR surrogate model to a subsurface pressure management problem with a random Gaussian permeability field. We solve the Darcy single-phase steady-state equation that examines the long-term impact of the injection/extraction on the reservoir. We consider that the pressure at the critical location is influenced strongly by the random permeability field during injection/extraction. Our GPR model predicts the pressure at a fixed critical location in the subsurface from an extraction rate and (truncated) permeability field realization. After we train the GPR surrogate model offline, it is used online to quickly determine the smallest extraction rate required to preserve a pressure at the critical location below a threshold with high probability.

Two discretizations must be made to solve the problem. First, we truncate the Karhunen-Loève expansion of the Gaussian permeability field to a finite sum. The random coefficients in this sum determining the (truncated) permeability field are inputs to the GPR model alongside the extraction rate. Second, the domain must be meshed in order to apply a finite volume method to solve the PDE numerically. Each of these discretizations induces a numerical error, and these errors are often neglected in subsurface flow problems. By contrast, our GPR modeling approach accounts for these errors in the uncertainty analysis.

Our novel contributions are as follows. First, we use ideas in multi-level Monte Carlo to derive an approximate upper bound on the root mean squared error between the discretized numerical solution and the analytic PDE solution. Then, this upper bound is used as an initial guess for the noise variance in our GPR model before hyperparameter optimization. Finally, we use a quasi-random design of experiments and matching covariance kernel to accelerate GPR model fitting and hyperparameter optimization from the classic $\mathcal{O}(n^3)$ rate to $\mathcal{O}(n \log n)$. 

These ideas enable error-aware GPR modeling that can scale to tens or even hundreds of thousands of observations for an accurate fit in high dimensions. Moreover, the GPR predictions come with a notion of uncertainty. In fact, the GPR surrogate is a distribution over possible functions mapping the extraction rate and permeability field to pressure at the critical location. 
%The GPR surrogate estimates the subsurface pressure field and also models its uncertainty by providing a probability distribution over a broad class of possible solutions. 

%\noindent 
In conclusion, we would like to summarize our findings:
\begin{itemize}
    \item The GPR fitting and optimization scales like $\mathcal{O}(n \log n)$ in the number of numerical PDE pressure solutions. %to fit. 
    \item The GPR model is error-aware by calibrating surrogate noise to numerical errors in solving the subsurface flow problem.
    \item The surrogate model quantifies the uncertainties in the predicted pressures by providing a probability distribution over a broad class of possible solutions. 
    \item In addition to subsurface flow problems, our approach can be directly applied to a variety of other problems that consider PDEs with random coefficients. 
\end{itemize}

Our primary focus here is solving problems in the context of pressure management to prevent overpressurization in the subsurface due to climate mitigation operations such as injecting wastewater or CO$_2$ sequestration. To allow for CO$_2$ sequestration applications, a more complex multiphase flow model would be needed, but the process for applying the GPR would remain the same. The key common ground is the existence of PDEs with random coefficients, which is very common in subsurface applications where the random coefficients are used to represent subsurface heterogeneity, e.g., in permeability fields in subsurface flow or velocity fields in seismic problems.
  
%We have fit a GPR surrogate to the random subsurface pressure at a critical location induced by a random Gaussian permeability field. Our method is scalable by using Gaussian processes that cost $\mathcal{O}(n \log n)$ to fit, error aware in calibrating noise in the Gaussian process to numerical errors in solving the PDE, and transferable to a variety of other problems. While this paper has only focused on Darcy's equation, we believe the presented methodology is almost immediately applicable to other PDEs with random coefficients. 

%\notes{In some disciplines, the use of Discussion or 'Conclusion' is interchangeable. It is not mandatory to use both. Please refer to Journal-level guidance for any specific requirements.}

\section*{Acknowledgements}
%\notes{Nick and Dan need to add acknowledgments.}

Aleksei Sorokin, Aleksandra Pachalieva, Nicolas Hengartner and James Hyman acknowledge the Center for Non-Linear Studies at Los Alamos National Laboratory. The work was supported by the U.S. Department of Energy through the Los Alamos National Laboratory. Los Alamos National Laboratory is operated by Triad National Security, LLC, for the National Nuclear Security Administration of U.S. Department of Energy (Contract No. 89233218CNA000001). Fred Hickernell and Aleksei Sorokin acknowledge support from the U.S. National Science Foundation grant DMS-2316011. NicolasHengartner acknowledge supported from LDRD grant 20210043DR.  The views expressed herein do not necessarily represent the views of the U.S. National Science Foundation, the U.S. Department of Energy, or the United States Government.

\bibliographystyle{elsarticle-num} 

\bibliography{main}

\end{document}